\documentstyle[prl,aps,floats,epsf,color,graphicx]{revtex}
\begin{document}
\baselineskip=12pt
\def\be{\begin{equation}}
\def\ee{\end{equation}}
\def\ba{\begin{eqnarray}}
\def\ea{\end{eqnarray}}
\def\la{\langle}
\def\ra{\rangle}
\def\a{\alpha}
\def\b{\beta}
\def\m{\mu}
\def\n{\nu}
\def\h{\hskip 1cm}
\def\hh{\hskip 2cm}
\def\lo{\longrightarrow}
\twocolumn[\hsize\textwidth\columnwidth\hsize\csname@twocolumnfalse\endcsname

\title { Exact symmetry breaking ground states for quantum spin chains}

\author {Sahar Alipour$^{a}$, Sima Baghbanzadeh,$^{b}$ and Vahid Karimipour$^{a,c}$}

\vskip 1cm

\address
{$^a$  Department of Physics, Sharif University of Technology, P.O.
Box 11365-9161, Tehran, Iran.}
\address
{$^b$  Department of Physics, Iran University of Science and
Technology, P.O. Box 16765-163, Tehran, Iran.}
\address
{$^c$  Corresponding author,Email:vahid@sharif.edu}\maketitle

%%%%%%%%%%%%%%%%%%%%%%%%%%%%%%%%%%%%%%%%%%%%%%%%%%%%%%
%ABSTRACT zittartz paper, many more references, so(3) and so(2), our own method, symmetry, etc.. reading more papers...
%maximum and minimum energy, projection of excited states on spin3/2,
%emphasis on the top and bottom states,
%%%%%%%%%%%%%%%%%%%%%%%%%%%%%%%%%%%%%%%%%%%%%%%%%%%%%%
\begin{abstract}
We introduce a family of spin-$1/2$ quantum chains, and show that
their exact ground states break the rotational and translational
symmetries of the original Hamiltonian. We also show how one can use
projection to construct a spin-$3/2$ quantum chain with nearest
neighbor interaction, whose exact ground states break the rotational
symmetry of the Hamiltonian. Correlation functions of both models
are determined in closed form. Although we confine ourselves to
examples, the method can
easily be adapted to encompass more general models. \\
PACS:  75.10.Jm
\end{abstract}

\newpage]

The question of exact ground state of a given quantum many body
system, i.e. a quantum spin chain, has always been fascinating and
of great importance to physicists in all fields, from condensed
matter physics and statistical mechanics to mathematical physics and
quantum field theory \cite{Baxter,rev1,rev2}. Besides their inherent
interest as a theoretical laboratory, where the phenomenon of
quantum phase transition \cite{sach} can be studied in detail and in
analytical form, quantum spin chains provide an important link with
classical statistical mechanics on two dimensional lattice models,
the latter is a subject with a rich and fascinating variety of
cooperative and critical phenomena affording insights into many
properties of real physical, chemical and biological systems.

As prototypes of correlated many-body systems, the Hilbert space
dimension of quantum spin chains grows exponentially with system
size, rendering any numerical treatment of large systems extremely
difficult. Therefore any exact solution of a new quantum spin chain,
is not only valuable for its own specific characteristics, but may
provide us with a plethora of mathematical techniques for
constructing a whole new series of exactly solvable models.

In this letter, we introduce a new model and technique for
constructing exact ground states of quantum spin chains which
spontaneously break the symmetries of the parent Hamiltonian. The
method is inspired by the original work of Affleck, Kennedy, Lieb
and Tasaki who first introduced it for constructing exact invariant
ground states for certain quantum chains \cite{aklt}. Their method,
designed in the context of valence bond solids, led in the course of
time to the formalism of finitely correlated or matrix product
states \cite{fcs}.  The model we introduce is a spin $1/2$ quantum
system on a ring, with interaction range equal to $4$, having both
translational and rotational symmetry, nevertheless we show that its
exact ground states break both these symmetries. We also find the
exact maximum energy state and some of the exact excited states, a
progress which has not been reported for any of the models
constructed in the AKLT or the matrix product formalism
\cite{mpsworks1,mpsworks2,AKM,KM}. Finally we use this construction
to introduce another quantum spin chain, namely a spin $3/2$ quantum
chain with nearest neighbor interaction, with symmetry breaking
ground states. We should stress that although we do this for a
specific model with rotational and translational symmetry, our
method can be adapted for construction of a broader class of similar
models.

Let us first fix our notation: we use $|+\ra$, $|-\ra$ and
$|S\ra:=\frac{1}{\sqrt{2}}(|+,-\ra-|-,+\ra)$ to denote respectively
the spin up, spin down and a singlet state of two spins. Subscripts
indicate the sites of the lattice, thus $|S_{_{12}},+_{_3}\ra$
indicates a singlet on sites 1 and 2 and a spin up state on site
$3$. Consider the state
\begin{equation}\label{state}
|\phi_0\ra:=|S_{_{12}},+_{_3},S_{_{45}},+_{_6},\cdots
S_{_{3N-2,3N-1}},+_{_{3N}}\ra,
\end{equation}
shown in figure (1).
 We require that $|\phi_0\ra$
be the exact ground state of a spin chain Hamiltonian
$H=\sum_{k=1}^{3N} h_{k,k+1,k+2,k+3}$, where we use periodic
boundary conditions. To this end, we demand that the local
Hamiltonian $h$ be a positive operator and annihilate $|\phi_0\ra$.
A local Hamiltonian $h$ finds four adjacent spins only in one of the
mixed states shown in figure (1). Thus, we should have
\begin{eqnarray}\label{BasicCondition}
% \nonumber to remove numbering (before each equation)
  &&tr\left[h(|S\ra\la S|\otimes |+\ra\la +|\otimes I)\right] = 0, \cr
  &&h\left[|+\ra|S\ra|+\ra\right]=0,\cr
  &&tr\left[h(I\otimes |+\ra\la +|\otimes |S\ra\la S|)\right] = 0.
\end{eqnarray}
Since $|S\ra$ is a singlet, the state $|+\ra|S\ra|+\ra$, can be
written as a linear combination of states with total spin $0$ and
$1$. Therefore a projector $P_2$ which projects the product of four
adjacent spins on the the sector with total spin $2$, automatically
satisfies equations (\ref{BasicCondition}). However there is one
other projector which satisfies equations (\ref{BasicCondition}),
and its annihilation of these states is quite nontrivial.  This is
the projector $P_0:=|\chi\ra\la \chi|$, where $|\chi\ra$ (or its
embedding on sites 1,2,3 and 4) is a singlet of four spins
constructed as follows:
\begin{equation}\label{chi}
    |\chi\ra=|S_{14},S_{23}\ra+|S_{13},S_{24}\ra.
\end{equation}
%%%%%%%%%%%%%%%%%%%%%%%%%%%%%%%%%%%%%%%%%%%%%%%%%%%%%%%%%%%%%%%%%%%%%%%%%
\begin{figure}
\epsfxsize=8.5truecm\epsfbox{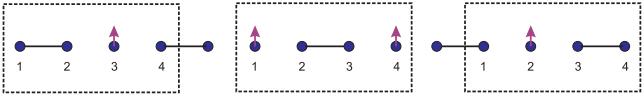} \narrowtext
\vspace{0.5cm}\caption{The three equations in (\ref{BasicCondition})
correspond, respectively from left to right, to the annihilation of
these local mixed states by $h$. For ease of comparison with the
analysis in the text we have labeled the spins in all boxes as 1, 2,
3 and 4.}\label{trimersANDH}
\end{figure}
To show this we use the simple facts $$\la
+_1|S_{12}\ra=\frac{1}{\sqrt{2}}|-_{2}\ra,\ \ \ \la
-_1|S_{12}\ra=\frac{-1}{\sqrt{2}}|+_{2}\ra,$$ to arrive at the
identity
\begin{equation}\label{identity}
    \la S_{23}|S_{12},S_{34}\ra=\frac{-1}{2}|S_{14}\ra,
\end{equation}
where $1,2,3$ and $4$ are any four different indices. From this
identity one can easily derive other identities simply by permuting
indices in an appropriate way, namely $\la
S_{34}|S_{14},S_{23}\ra=\frac{-1}{2}|S_{12}\ra,$ or $\la
S_{34}|S_{13},S_{24}\ra=\frac{1}{2}|S_{12}\ra$. These two
identities, immediately imply that $\la S_{34}|\chi\ra=0$. Similarly
one finds $\la S_{12}|\chi\ra =0$, hence the first and the last
equations of (\ref{BasicCondition}) are satisfied. For the second
relation of equation (\ref{BasicCondition}) we note that
\begin{equation}\label{last}
    \la \chi|S_{23}\ra=\la S_{14},S_{23}|S_{23}\ra+\la
    S_{13},S_{24}|S_{23}\ra=\frac{3}{2}\la S_{14}|,
\end{equation}
hence
\begin{equation}\label{3}
\la \chi|+_1,S_{23},+_4\ra=\frac{3}{2}\la S_{14}|+_1,+_4\ra=0.
\end{equation}
This proves the assertion that $P_0$ annihilates all three kinds of
local states. Thus, we find that the local Hamiltonian can be of the
form $h=2JP_2+P_0$, where $J$ is a positive but arbitrary coupling
constant, the factor of 2 is for convenience and one of the
couplings, the coefficient of $P_0$, has been re-scaled to unity.
The final form of the Hamiltonian in terms of spin operators can be
found as follows. First we note that $\frac{1}{2}^{\otimes
4}=2\oplus 1^{3}\oplus 0^{2}$ where the exponents indicate the
multiplicities and $P_0$ is the projector to one of the above two
zeros. Let us denote the sum of spin-$j$ projectors in the above
decomposition by ${\cal P}_j$. Then, from ${S\cdot
S}=\sum_{j}j(j+1){\cal P}_j$, where $S$ is the spin operator of four
consecutive sites, we have
\begin{eqnarray}\label{projectoreqauations}
% \nonumber to remove numbering (before each equation)
  I &=& {\cal P}_0+{\cal P}_1+{\cal P}_2, \cr
  S\cdot S &=& 2{\cal P}_1+6{\cal P}_2, \cr
  (S\cdot S)^2 &=& 4{\cal P}_1+36{\cal P}_2,
\end{eqnarray}
from which we obtain
\begin{eqnarray}\label{projectors}
% \nonumber to remove numbering (before each equation)
  {\cal P}_0 &=& 1-\frac{2}{3}{S}\cdot {S}+\frac{1}{12}({S}\cdot{S})^2, \cr
  {\cal P}_2 &=& -\frac{1}{12}{S}\cdot {S}+\frac{1}{24}({S}\cdot{S})^2.
\end{eqnarray}
This readily gives ${P}_2=({\cal P}_2)$ in terms of spin operators.
However, the individual projector $P_0$ can not be obtained in this
way. Direct calculation from (\ref{chi}) is notoriously tedious. We
can circumvent this problem by calculating the other projector
$P'_0=|\chi^{\perp}\ra\la \chi^{\perp}|$ and subtract it from ${\cal
P}_0$ to obtain $P_0$. It is straightforward to check from
(\ref{identity}) that the other singlet which should be
perpendicular to $|\chi\ra$ is the following
\begin{equation}\label{chiperp}
    |\chi^{\perp}\ra=|S_{12},S_{34}\ra.
\end{equation}
This readily gives $$ P_0={\cal P}_0 - |\chi^{\perp}\ra\la
\chi^{\perp}\ra={\cal P}_0 - (\frac{1}{4}-{\bf s}_1\cdot {\bf
s}_2)(\frac{1}{4}-{\bf s}_3\cdot {\bf s}_4). $$ Putting all these
together, and discarding overal constants, we find
\begin{eqnarray}\label{norm}
H=\sum_{i=1}^{3N} && {\bf s}_i\cdot {\bf s}_{i+1}(1-2{\bf
s}_{i+2}\cdot {\bf s}_{i+3})\cr&-&\frac{4+J}{3}{\bf S}_i\cdot {\bf
S}_{i}+\frac{1+J}{6}({\bf S}_i\cdot {\bf S}_i)^2,
\end{eqnarray}
where $s_i$ is the spin operator of one site and ${\bf S}_i$ is the
spin operator of a block of four spins at
sites $i,i+1,i+2,i+3$.
Note that the relative strength of consecutive couplings are approximately 1: 0.48: 0.06.\\

The state $|\phi_0\ra$ is the exact ground state of $H$ which breaks
translational and rotational symmetry of the Hamiltonian. Clearly,
it is not invariant under rotation, it is the top state of the total
spin-$\frac{N}{2}$ sector, other states which are degenerate with it
in energy, are obtained by acting with total
$L_-=\sum_{k=1}^{3N}\sigma^-_{k}$ operator, which does not affects
the singlets but flips the up spins. None of these $N+1$ states, are
invariant under translation. Thus the ground state has a $3(N+1)$
fold degeneracy, generated by the broken symmetries of the
Hamiltonian, namely the continuous $so(3)$ symmetry and the discrete
translational symmetry. To find translational invariant states, we
form the states $|\Psi^j\ra$, $j=0,1,2$ as follows:
%%%%%%%%%%%%%%%%%%%%%%%%%%%%%%%%%%%%%%%%%%%%%%%%%%%%%%%%%%%%%%%%%%%%%%%%%
\begin{figure}
\epsfxsize=7.5truecm\epsfbox{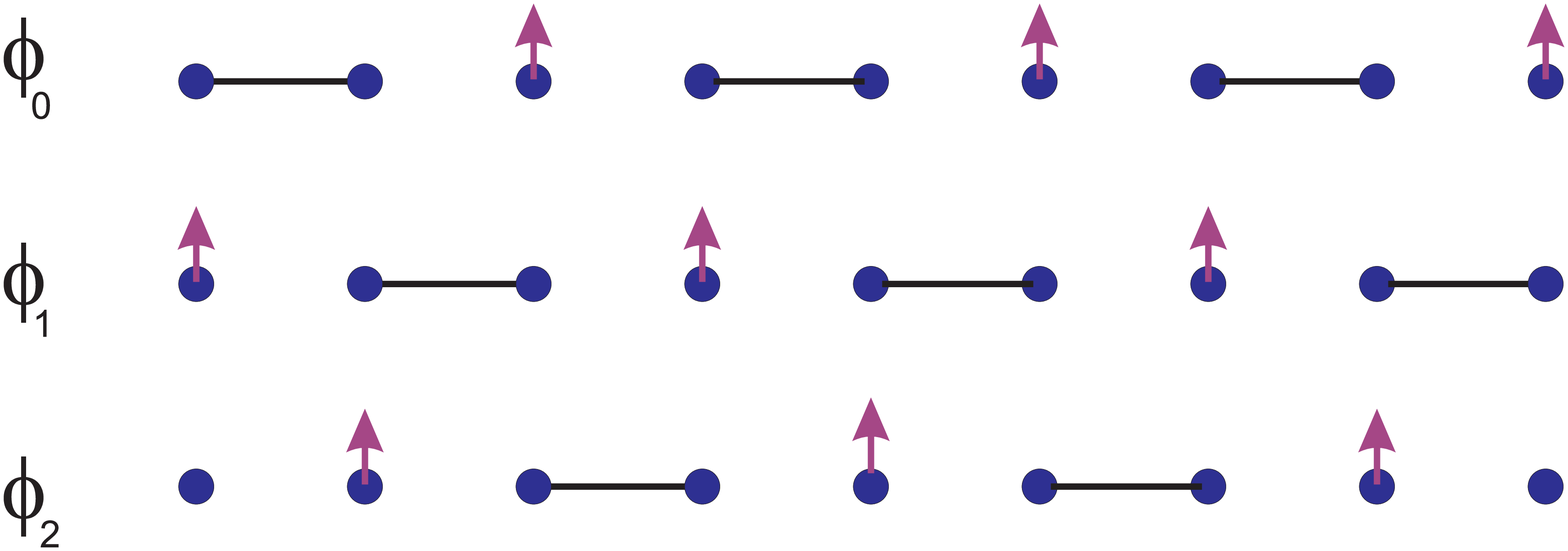} \narrowtext
\vspace{0.5cm}\caption{The states $|\phi_0\ra$, $|\phi_1\ra$ and
$|\phi_2\ra$.}\label{trimers}
\end{figure}
\begin{equation}\label{translation}
    |\Psi^j\ra:=\frac{1}{\sqrt{3}}(|\phi_0\ra + \omega^j\ |\phi_1\ra
    +\omega^{2j}\ |\phi_2\ra),
\end{equation}
where $\omega^3=1$, and $|\phi_i\ra=T|\phi_{i-1}\ra$, where $T$ is
the one-site translation operator, figure (2). These states are
eigenstates of $T$, with eigenvalues $1,\omega$ and $\omega^2$
respectively. One can readily show that for $i\ne j$,
$$\la \phi_i|\phi_j\ra = (\la +,S|S,+\ra)^N=(\frac{-1}{2})^N,$$ from
which we find the states $|\Psi^j\ra$ are normalized in the
thermodynamic limit. In this same limit, the correlation functions
are easy to find. First the magnetization turns out to be
\begin{equation}\label{mag}
    \la \Psi^j|S^{z}|\Psi^j\ra=\frac{1}{6},\ \ \ \forall\  j.
\end{equation}
and the two-point correlation function becomes
\begin{eqnarray}\label{cor}
    \la \Psi^{j}|S^{z}_1 S^{z}_{r}|\Psi^j\ra &=&\frac{-1}{12}
    (\delta_{r,2}-\sum_{n=1}\delta_{r,{3n+1}}),\ \ \ \
    \cr
    \la \Psi^{j}|S^{x}_1 S^{x}_{r}|\Psi^j\ra &=&\frac{-1}{12}\delta_{r,2},\h \ \ \
    \forall j.
\end{eqnarray}
%%%%%%%%%%%%%%%%%%%%%%%%%%%%%%%%%%%%%%%%%%%%%%%%%%%%%%%%%%%%%%%%%%%%%%%%%
We now ask if it is possible to find some of the exact excited
states of the Hamiltonian. We will show that this is indeed
possible, although the states thus found are not the low-lying
states of the energy spectrum. To proceed, we note that the state
\begin{equation}\label{Omega}
    |\Omega\ra:=|+,+,+,\cdots,+,+\ra,
\end{equation}
is an exact energy eigenstate, indeed the maximum energy eigenstate
with $E_{max}=6NJ$ (here we are working with $h=2JP_2+P_0$ and not
with (\ref{norm}), in which multiplicative and additive constants
have been ignored). To see this note that $P_0|\Omega\ra=0$ and
$P_2|\Omega\ra=|\Omega\ra$. The maximality of the energy is found by
taking $J$ very large and noting the continuity of the energy with
respect to $J$. Consider now the one ''hole'' state of the form
$$|n\ra = |+,+,\cdots, - ,\cdots , +, +\ra,$$ where the state in the $n-th$ position has been flipped.
A general one-hole state is of the form
\begin{equation}\label{Psiexited}
    |\psi^r\ra=\sum_{n=1}^{3N}\psi^{r}_n|n\ra.
\end{equation}
It is obvious that $P_0|n\ra=0$, for any embedding of projection
$P_0$ on four consecutive sites. To see this, note that the total
$z$ component of each four consecutive spins is either 2 or 1.  Also
from the explicit form of the spin-2 states, one can check that
$P_2$ does not take a state $|n\ra$ outside the one-hole sector. In
fact, a simple calculation shows that
$$
    H|n\ra = 6J(N-1)|n\ra + \frac{J}{2}\sum_{k=1}^3 (4-k)(|n+k\ra + |n-k\ra).
$$ Inserting this in the eigenvalue equation
$H|\psi^r\ra=E_r|\psi^r\ra$, gives
\begin{equation}\label{Psiexited}
    |\psi^r\ra=\sum_{n=1}^{3N}e^{in\frac{2\pi r}{3N}}|n\ra,
\end{equation}
with
\begin{equation}\label{HPsi}
     E_r = 2J(3N-4+2\cos^3 \frac{2\pi
    r}{3N}+2\cos^2 \frac{2\pi r}{3N}).
\end{equation}
Note that $-E_{max}$ is the exact ground state energy of $-H$, in
which case $-E_r$ gives the energy of the low-lying excitations or
spin waves of $-H$. The dispersion relation is plotted in figure
(3).
\begin{figure} \centering\epsfxsize=5truecm\epsfbox{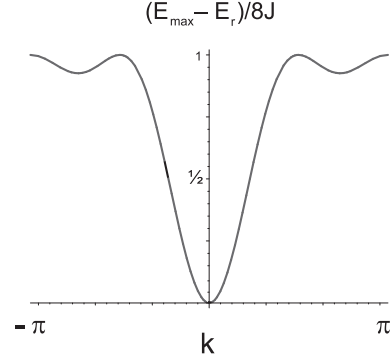}
\narrowtext\vspace{0.5cm} \caption{The dispersion relation for the
holes of model (\ref{norm}).}\label{trimersProjection}
\end{figure}
Note that if it were not for the $P_0$ projector, then a
Majumdar-Ghosh state, a singlet, would have been the ground state of
$H$. The role of $P_0$ is to elevate the energy of this state and
hence break the symmetry. Finally we note that from the above model
one can obtain an exact ground state for a spin $\frac{3}{2}$ model
with nearest neighbor interaction. Let us project three spins into a
spin $\frac{3}{2}$ site as shown in figure (4).
\begin{figure}
\epsfxsize=8truecm\epsfbox{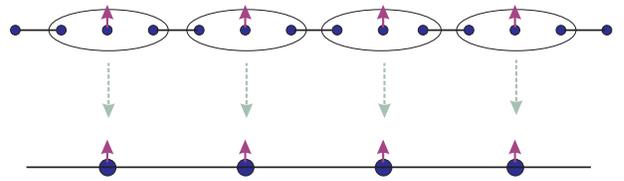}
\narrowtext\vspace{0.5cm} \caption{The states in each bulb, are
projected to a spin $3/2$ state, to form a new quantum spin chain
with spin equal to $3/2$ and with nearest neighbor
interactions.}\label{trimersProjection}
\end{figure}
The product of two
adjacent spin $\frac{3}{2}$'s on the lower chain, being effectively
the product of four spin $\frac{1}{2}$'s on the upper chain, can
combine to give only spins $2$, $1$ and $0$. Therefore a projector
$P_3$ on spin $3$ sector, can annihilate the local state of any two
adjacent spins in the lower chain. This means that the state
constructed as such, will be an exact ground state of a Hamiltonian
with nearest neighbor interaction $H=\sum_{k}({P_3})_{_{k,k+1}}$.
The projector is found in the same way as in
(\ref{projectoreqauations}) and (\ref{projectors}) and we arrive at
$$
    H=C+\sum_{i=1}^N \frac{81}{8}{\bf S}_i\cdot {\bf S}_{i+1}+\frac{29}{6}({\bf S}_i\cdot {\bf S}_{i+1})^2
    +\frac{2}{3}({\bf S}_i\cdot {\bf S}_{i+1})^3,
$$ where $C=\frac{165}{32}N$. In this way, we construct
a rotationally invariant Hamiltonian whose exact ground state breaks
this rotational symmetry. The remarkable point is that we can also
obtain the explicit form of this state, beyond the construction
shown in figure (4). With this explicit expression one can then
determine the correlations in this state which are no longer
restricted to nearest neighbor sites. To achieve this, let us first
construct a Matrix Product (MP) representation for the parent
(upper) dimer state in figure (4). In view of the MP representation
\cite{KM} of dimerized Majumdar-Ghosh state \cite{MG}, this state
has the following representation:
$$|\phi_0\ra=tr(A_{_{i_1}}A_{_{i_2}}B_{_{i_3}}A_{_{i_4}}A_{_{i_5}}B_{_{i_6}}\cdots)|i_1,i_2,i_3,i_4,i_5,i_6,\cdots\ra
$$ where
\begin{eqnarray}\label{matrices}
   A_{_+}&=&|+\ra\la 0|+|0\ra\la -|,\cr
    A_{_-}&=&|-\ra\la 0|-|0\ra\la +|,\cr
    B_{_+}&=&|0\ra\la 0|, \ \ \ B_-=0,
\end{eqnarray}
and $|+\ra,|-\ra$ and $|0\ra$ are three ortho-normal states. Using
the Clebsh-Gordon coefficients to decompose the product of three
states in a bulb of the upper chain in figure (4) and noting that
$B_{_-}=0$, we find the following matrices corresponding to the
states of a single spin $3/2$ site on the lower chain:
\begin{eqnarray}\label{matrices}\nonumber
    A_{_{3/2}}&=&A_{_+}B_{_+}A_{_+},\ \ \ A_{_{1/2}}=\frac{1}{\sqrt{3}}(A_{_-}B_{_+}A_{_+}+A_{_+}B_{_+}A_{_-}),\cr
    A_{_{-3/2}}&=&0, \ \ A_{_{-1/2}}=\frac{1}{\sqrt{3}}A_{_-}B_{_+}A_{_-},
\end{eqnarray}
leading to
$$ A_{_{3/2}}=\sigma^+,\ \ A_{_{1/2}} = \frac{-1}{\sqrt{3}}\sigma_z,\ \ A_{_{-1/2}}= \frac{-1}{\sqrt{3}}\sigma_-,\ \
    A_{_{-3/2}}=0. $$
 The state of the lower chain is now
given by a matrix product state, with the above matrices. The ground
state is a multiplet with total spin $N/2$, for which we have
depicted the top state explicitly. This MP representation for spin
3/2 chain has already been obtained in \cite{zit32} by other
methods. Having a MP representation, it is straightforward to
calculate the correlation functions. The results which agree with
that of \cite{zit32} are that $\la S^z\ra=\frac{1}{2},\ \ \la
S^x\ra=\la S^y\ra=0,$ and
$$ \la ({S^z})^2\ra= 5/4-u,\ \ \ \ \la(S^x)^2\ra=\la
(S^y)^2\ra=5/4+\frac{u}{2},
$$
where $u:=\frac{1}{1+\sqrt{3}}$. With the notation $G^a(1,r):=\la
S^a_1S^a_r\ra-\la S^a_1\ra\la S^a_r\ra$,
$$     G^z(1,r)=A_l (-1)^{r-1}\ e^{-\frac{r}{\xi_l}}\ ,\ \  G^t(1,r)=A_t(-1)^{r-1}\ e^{-\frac{r}{\xi_t}}
$$ where
\begin{equation}
    A_l:=\frac{12+6\sqrt{3}}{4},\ \ \ \ A_t:=\frac{12+7\sqrt{3}}{4},
\end{equation}
and
\begin{equation}\label{correlationlengths}
    \xi_l=\frac{1}{\ln(2+\sqrt{3})},\ \ \ \ \xi_t=\frac{1}{\ln(1+\sqrt{3})}.
\end{equation}
The correlation functions for other states in the ground state
multiplet can be determined by the application of rotational
symmetry. In summary, we have introduced a one-parameter family of
isotropic spin 1/2 quantum chains, whose exact ground states break
the translational and rotational symmetry of the Hamiltonian. The
totality of degenerate vacua of these models are generated by the
application of rotation and translation operators. Besides the
ground state, the maximum energy state and some of the lower energy
states have also been derived in closed analytical form. A related
spin $3/2$ quantum chain with nearest neighbor interactions has also
been constructed, the exact ground state of which has an MP
representation. The method presented in this letter can be pursued
further, to make other spin models whose ground state break the
rotational symmetry of the Hamiltonian. There is also an interesting
un-answered question which deserve further investigation: Is there a
singlet state which is invariant under rotation and has a higher
energy than the degenerate ground states \cite{singlet}? V. K. would
like to thank Laleh Memarzadeh for very valuable and constructive
discussions. S. A. and S.B. would also thank Ali Rezakhani for
valuable discussions.
{}

\begin{thebibliography}{99}
\bibitem{singlet} Direct and lengthy calculation shows that for a ring of size $4$,
 $|\a\ra:=|S_{_{14}},S_{_{23}}\ra-|S_{_{12}},S_{_{34}}\ra$ and $|\beta\ra:=|S_{_{13}},S_{_{24}}\ra$ are eigenstates with
  energies 1 and 3 resp. and for a ring of size $6$,
$|\gamma\ra=|S_{_{16}},S_{_{25}},S_{_{34}}\ra-2|S_{_{12}},S_{_{36}},S_{_{45}}\ra+|S_{_{14}},
S_{_{23}},S_{_{56}}\ra$ is an eigenstate with energy $2$.
\end{thebibliography}
\end{document}